\documentclass[
	pra,
	aps,
	reprint,
	amsmath,amssymb,
	a4paper,
	superscriptaddress,
	10pt,
    nofootinbib
]{revtex4-2}

\usepackage{times}      %
\usepackage{helvet}     %
\usepackage{physics}    %
\usepackage{upgreek}    %

\usepackage{graphicx}
\graphicspath{{Fig/}}
\usepackage[dvipsnames]{xcolor}
\definecolor{rmpblue}{HTML}{2e3092}

\usepackage{enumitem}

\newcommand{\nm}{\ensuremath{\,\text{nm}}}            %
\newcommand{\gtwo}{g^{(2)}}          %
\newcommand{\hBN}{h-BN}              %
\newcommand{\SPE}{SPE}               %

\newcommand{\affilBombay}{Laboratory of Optics of Quantum Materials, Department of Physics,
Indian Institute of Technology Bombay, Mumbai 400076, India}

\begin{document}
\title{Deterministic Single-Photon Emitter Arrays in Hexagonal Boron Nitride by Carbon-Assisted Focused Ion Beam Engineering}

\author{Mangababu Akkanaboina}
\thanks{Equal author contribution}
\affiliation{\affilBombay}
\author{Rohit Kumar}
\thanks{Equal author contribution}
\affiliation{\affilBombay}
\thanks{Equal author contribution}
\author{Brijesh Kumar}
\thanks{Equal author contribution}
\affiliation{\affilBombay}
\author{Hrushikesh Gawali}
\affiliation{\affilBombay}

\author{Parul Sharma}
\affiliation{\affilBombay}

\author{Ikshvaku Shyam}
\affiliation{\affilBombay}

\author{Anshuman Kumar}
\email{anshuman.kumar@iitb.ac.in}
\affiliation{\affilBombay}

\begin{abstract}
The realization of on-chip photonic circuits requires scalable and deterministic single-photon emitters (\SPE{}s) at room temperature, which remain a challenge in van der Waals materials. In this work, we report a novel three-step fabrication process for the generation of spatially controlled \SPE{} arrays in hexagonal boron nitride (\hBN{}). The process comprises site-selective gallium (Ga) focused ion beam milling, nanoscale conformal carbon deposition over the patterned regions, and subsequent thermal annealing. The synergistic combination of these steps resulted in a site-correlated emitter yield of ($\sim 89\%$) across 100 fabrication sites.  Second-order autocorrelation measurements revealed pronounced three-level emitter dynamics where the best emitters exhibited high purity ($\gtwo(\tau=0)=0.15 \pm 0.09$).To the best of our knowledge, this is the first lithography-free, direct-write approach combining Ga-ion milling, selective carbon engineering, and thermal annealing to deterministically generate \hBN{} \SPE{}s. The reproducibility of the method is validated across multiple independently fabricated samples.  These results establish a scalable, lithography-free pathway toward on-demand \SPE{} arrays relevant to integrated quantum photonics.

\medskip\noindent
\textbf{Keywords:} Single-photon emitter, Two-dimensional materials, Quantum photonics
\end{abstract}
\maketitle

\section{Introduction}
Realization of scalable quantum photonic architectures for applications in quantum technology such as quantum key distribution, optical quantum computing~\cite{Vogl2026,Scarani2009,Mehic2020,Anuj_prospective}, quantum sensing~\cite{Degen2017,Vaidya2023,Rizzato2023}, and super resolution imaging requires stable room temperature operable deterministic quantum emitter arrays~\cite{Hou2025}. Solid state quantum emitters in van der Waals materials are the highly promising candidates in terms of the ease of integration, near-field coupling to the photonic circuits, and their scalability \cite{Aharonovich2016,Daggett2026,Singh2025,aharonovich2016solid}. Hexagonal boron nitride (hBN) is well known for its blinking free, stable quantum emission at room temperature with the added advantage of 2D nature to integrate into photonic circuits \cite{Singh2025}. Hence, researchers across the globle are currently exploring the deterministic ways to create SPEs in hBN matrix using various techniques such as electron irradiation, AFM based strain engineering, laser irradiation, ion irradiation, controlled etching and  nano-indentation. For example, Kumar et al. \cite{Kumar2023} reported the electron irradiation of hBN flakes for the deterministic creation of SPEs with yellow emission (575 nm). Ahmed et al. \cite{Ahmed2025} have studied the role of nanoindentation and Tran, Toan Trong, et al.\cite{tran2016robust}  have studied thermal annealing on the creation of hBN SPEs. Luo et al. \cite{Luo2025} has explored the carbon-functionalization of hBN using micro-sphere lithography combined with ultrasonic nano-indentation for the creation of deterministic SPEs at 560 nm and achieved a yield of 59\%. 
\\Although significant progress has been made toward deterministic SPE generation in hBN, many existing approaches depend on lithography-assisted patterning and multiple processing steps, offer limited control over defect chemistry, or achieve only moderate activation yields under deterministic fabrication conditions. The origin of quantum emission reported in these studies, as well as in density functional theory calculations, has been widely attributed to the formation of carbon-related defect states, particularly $C_{B}V_{N}$ antisites and C--N complexes \cite{Park2026, Sajid2018}. These defect configurations are predicted to give rise to bright visible emission in the 580--640 nm spectral range. Nevertheless, the precise microscopic origin of the emitting centers remains an active area of investigation and has yet to be conclusively established. In this context, the deterministic incorporation and engineering of carbon-related defects within the hBN lattice, particularly through lithography-free fabrication approaches, may provide a promising route toward the scalable realization of spatially controlled SPE arrays.
\begin{figure*}
    \centering
    \includegraphics[width=\linewidth]{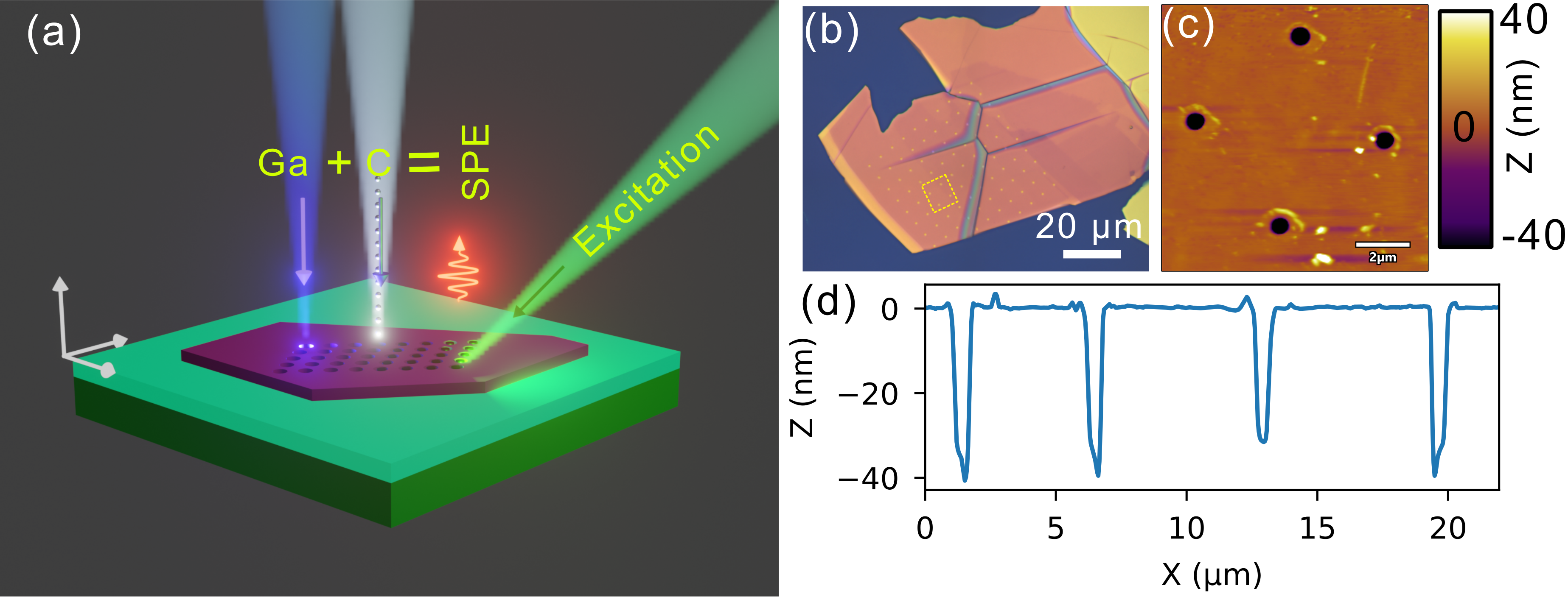}
    \caption{\textbf{a)} Schematic representation of sample preparation where the first step was milling the mechanically exfoliated hBN flake of thickness 120 \nm\ using Ga$^+$-ion based FIB, followed by selective carbon deposition at the milled spot and subsequent annealing at 850 $^\circ$C in an Ar environment. \textbf{b)} Optical microscope image of the processed hBN flake. \textbf{c)} AFM topography of the FIB-milled region marked by yellow rectangle in (b). \textbf{d)} Corresponding AFM height profile extracted across the milled spots, revealing a milling depth of $\sim 40 \nm.$}
    \label{fig1:sample_prepration}
\end{figure*}
\\ Focused-ion-beam (FIB) processing has recently attracted interest as a maskless, direct-write technique for engineering defects in 2D materials~\cite{eswaramoorthy2025protected,eswaramoorthy2026focused,Grosso2017}. Ga FIB milling can create sub-100 nm features with high positional accuracy and has been demonstrated to generate emitters in h-BN through displacement damage (Vacancy defects) \cite{Frch2021}. However, Ga-induced emitters typically lack spectral sharpness and cannot be selectively assigned to specific defect families \cite{Venturi2024}. Supplementing FIB milling with a targeted dopant species—such as carbon, which is closely linked to the sharp visible emitters in h-BN—offers a route to combine positional accuracy with chemical selectivity. In this study, we developed a novel 3 step fabrication flow for the creation of deterministic hBN emitter arrays. Ga milling is employed for fabrication of deterministic nano-groove array in the hBN matrix, subsequently a selective carbon deposition is carried out in the nano-grooves. Further, hBN flakes with nano-grooves filled with carbon is subjected to the thermal annealing at 850 $^\circ$C to realize the deterministic sharp SPE sources as shown in schematic Fig.~\ref{fig1:sample_prepration}(a). The combination of these three fabrication steps resulted in luminescent emission from $\sim 89\%$ of the targeted sites. The resulting emitters exhibit bright, photostable single-photon emission, demonstrating the effectiveness of the proposed approach for the deterministic generation of quantum emitters in hBN.
\begin{figure*}
    \centering
    \includegraphics[width = 0.9\linewidth]{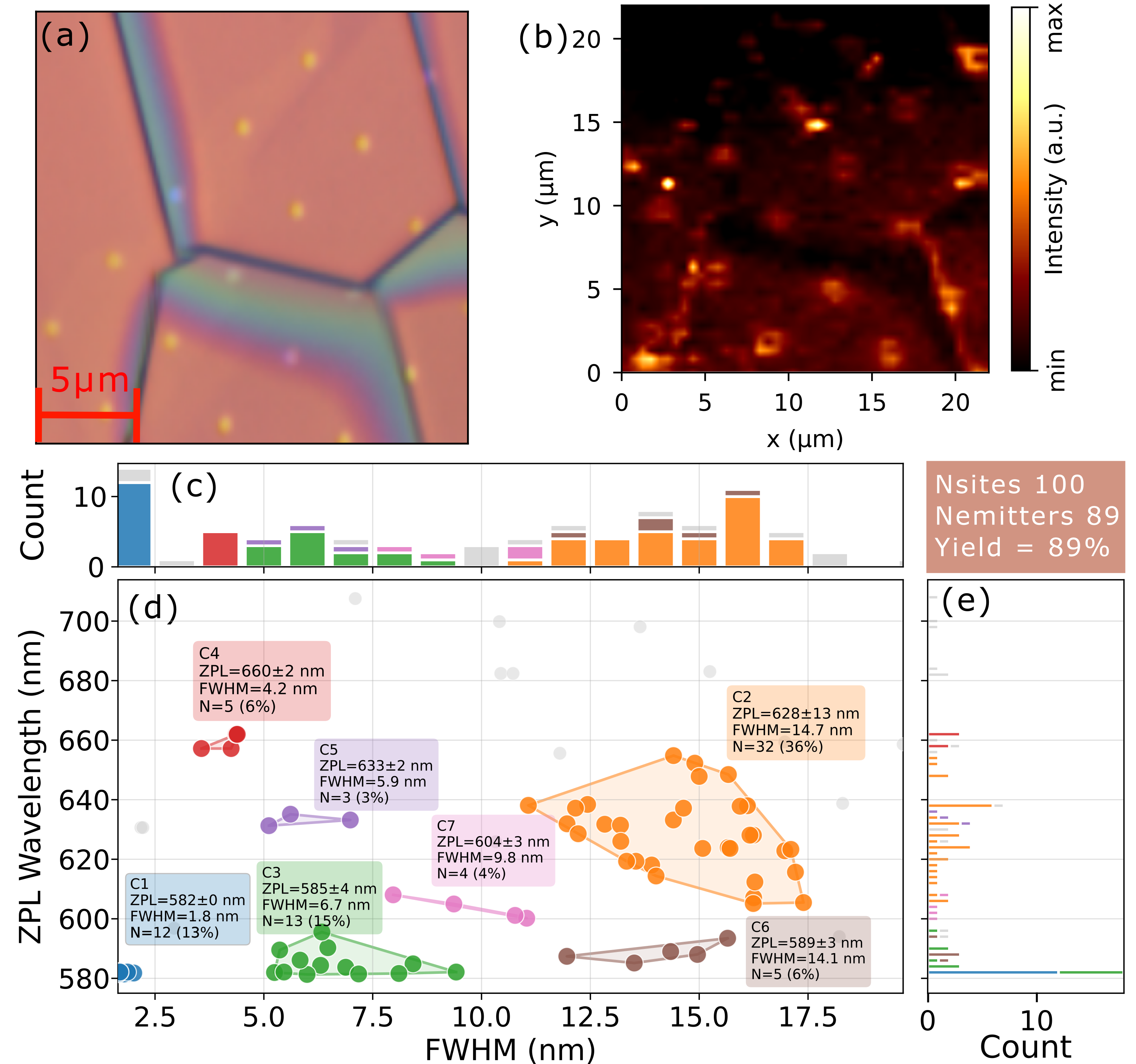}
    \caption{(a) Optical image of the milled spots where PL map is performed, (b) its corresponding PL map generated at a wavelength of 581 nm depicting luminescent array of milled spots. Statistical analysis of all $10 \times 10$ milled spots is shown in (c), (d), and (e). The scatter plot in (d) displays ZPL wavelengths of 89 emitters found with in $10 \times 10$ milled spots on Y-axis and their corresponding FWHM on X-axis, (C) displays the histogram of counts (number of emitters) vs FWHM, and (e) histogram of number of emitters vs wavelength}   
    \label{Fig2:statistical}
\end{figure*}
\section{Results and Discussion}
The fabricated deterministic emitter arrays on hBN flakes are clearly visible in optical image shown in Figure~\ref{fig1:sample_prepration}(b). Atomic force microscopy (AFM) analysis revealed the presence of nano-grooves  with a depth of $\sim 40 \nm\ $ as shown in the depth profile ~\ref{fig1:sample_prepration}(c) and  ~\ref{fig1:sample_prepration}(d). The confocal PL maps are conducted throughout the array in multiple $20 \times 20~\mu\mathrm{m}^2$ areal scans. Figure ~\ref{Fig2:statistical}(a) shows the optical image where one of the PL maps is performed, the corresponding PL map is shown in Figure ~\ref{Fig2:statistical}(b) confirms the presence of bright emission spots at the processed sites. The white circles in ~\ref{Fig2:statistical}(b) denote the milled sites. A detailed statistical analysis of the PL maps revealed that approximately $\sim 89\%$ of the fabricated sites exhibited sharp ZPL emission either within the milled spots or in the immediate vicinity from a $10 \times 10$ array. Figure~\ref{Fig2:statistical}(d) presents a scatter plot of the ZPL wavelength as a function of the corresponding full width at half maximum (FWHM) for all observed emitters.

The emitters were grouped into distinct spectral clusters based on their emission wavelength and linewidth. The most prominent emitter family is centered around 581--585 nm and comprises two closely related clusters: C1, centered at 581 nm with an average linewidth of 1.8 nm, and C3, centered at ($585 \pm 4$) nm with an average linewidth of 6.7 nm. These clusters account for approximately 13\% and 15\% of the total emitter population, respectively. Owing to their similar spectral positions, we consider C1 and C3 to belong to the same emitter family, which collectively represents ($\sim 28 \%$) of all observed emitters, making it the dominant spectral population in our dataset. Additional emitter clusters are identified at $604 \pm 3$ nm (C7), $632 \pm 2$ nm (C5), $640 \pm 3$ nm (C2), $660 \pm 2$ nm (C4), and $589 \pm 3$ nm (C6), with average linewidths ranging from approximately 4.7 to 14.1 nm. The histogram in Figure~\ref{Fig2:statistical}(c) summarizes the distribution of emitter counts as a function of FWHM, while Figure~\ref{Fig2:statistical}(e) shows the corresponding wavelength distribution of all detected emitters. The wavelength histogram reveals a clear predominance of emitters in $581$--$590$ nm spectral range, indicating a strong control over the formation of emitters with similar ZPL under the present fabrication conditions.  

The ZPL emission energies observed in this fabrication process fall within the range of 1.9-2.1 eV, strongly suggesting the formation of carbon complex defects \cite{Sajid2018, Tang2025}. It is well known that carbon can diffuse into the defect sites/vacancies in hBN lattice at higher annealing temperatures. Hence, in our case it is most probable that the Ga ion milling could produce a lot of vacancy related defects, selective deposition of carbon in the milled spots and thermal annealing at 850 $^\circ$C has led to formation carbon complex defects in the milled spots or in its vicinity.

A fraction of the observed emitters are located in the immediate vicinity of, rather than directly within, the FIB-milled sites. We attribute this behavior to thermally activated migration and reconfiguration of carbon- and vacancy-related defects during the annealing process, as reported elsewhere\cite{CHAKRABORTY2025}. Within the framework of classical diffusion theory, the diffusion coefficient is described by an Arrhenius relation\cite{Zobelli2007},
\begin{equation}
D = D_{0}\exp\left(-\frac{Q}{k_{B}T}\right),
\label{eq}
\end{equation}
where ($D_{0}$) is the diffusion prefactor, (Q) is the migration barrier, ($k_{B}$) is the Boltzmann constant, and (T) is the absolute temperature. The prefactor can be estimated from transition-state theory as ($D_{0}\propto a^{2}\nu_{0}$), where (a) is the atomic jump distance and ($\nu_{0}$) is a characteristic vibrational frequency of the lattice. Previous first-principles studies have shown that carbon-related defects and vacancies in hBN possess thermally accessible migration pathways under elevated-temperature annealing conditions, enabling defect redistribution and complex formation within the vicinity of the targeted sites~\cite{Weston2018, Zobelli2007}. Although the precise diffusion mechanism and migration barriers remain dependent on the specific defect configuration, the Arrhenius framework indicates that thermally activated defect migration is physically plausible under the annealing conditions employed in this work. Consequently, emitters observed slightly away from the nominal milled locations can still be regarded as spatially correlated with the targeted fabrication sites.
\begin{figure*}
    \centering
    \includegraphics[width=0.8\linewidth]{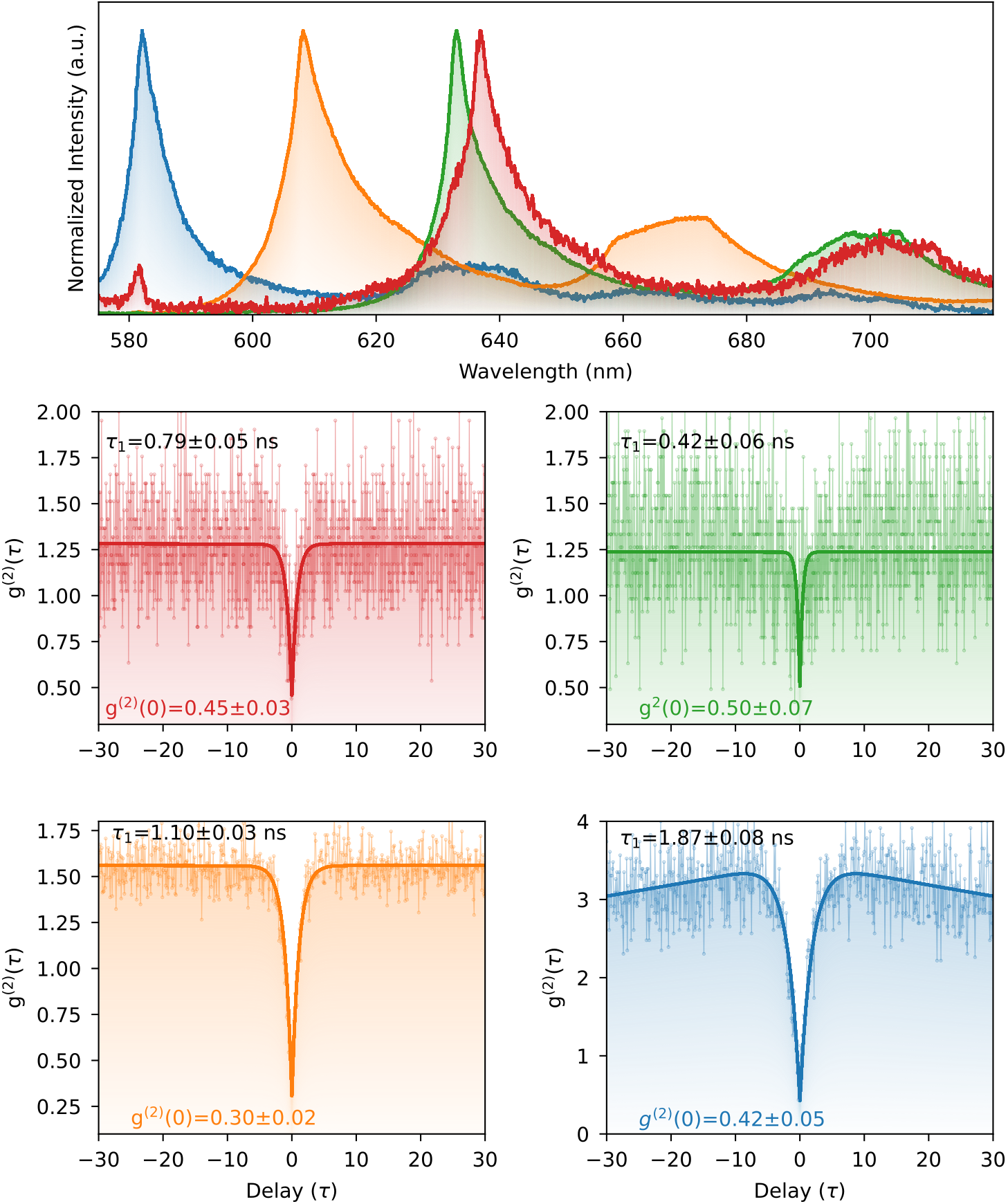}
   \caption{\textbf{Confirmation of quantum emitters.} \textbf{(a)} PL spectra of four different emitters identified in the mapped region.  \textbf{(b-e)}  Second-order autocorrelation measurements, $g^{(2)}(\tau)$, of the corresponding emitters shown in (a), confirming their single-photon emission characteristics.}
    \label{fig3:PL_g2}
\end{figure*}

\begin{figure*}
    \centering
    \includegraphics[width=0.95\linewidth]{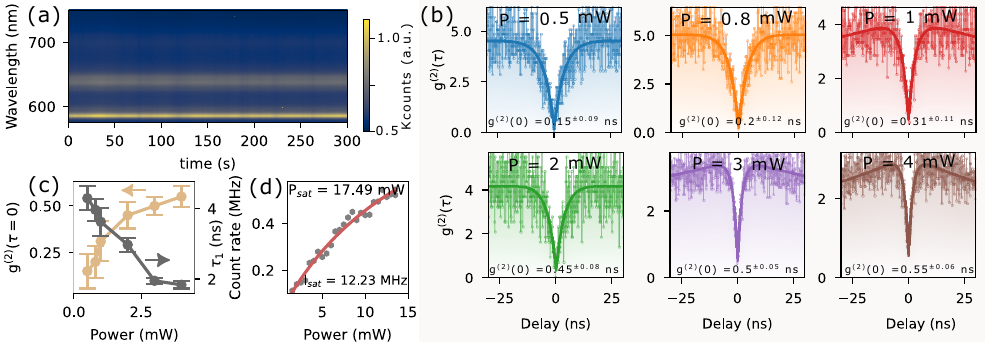}
   \caption{\textbf{Optical stability and performance metrics of emitters:} \textbf{(a)} Temporal stability measurement under continuous excitation, recorded over 5 min with a time bin of 0.5 s. \textbf{(b)} Second-order correlation $ g^{(2)}(0) $ as a function of excitation power ranging from 0.5 to 4 mW, confirming single-photon emission behavior.  \textbf{(c)} Second-order correlation $\gtwo(0)$ and characteristic antibunching time as a function of power. \textbf{(d)}   
   Power-dependent photoluminescence saturation curve of the emitter. }
    \label{fig4:stability}
\end{figure*}
Further, prominent ZPL emissions centered at 637, 633, 608, and 581 nm, identified through statistical analysis, are presented in Figure~\ref{fig3:PL_g2}(a). To verify the quantum nature of these representative emitters, detailed optical characterization was performed using a confocal photoluminescence (PL) setup integrated with a Hanbury Brown and Twiss (HBT) interferometer (see Methodology section for details). The corresponding room-temperature second-order autocorrelation measurements, $g^{(2)}(\tau)$ are shown in Figure~\ref{fig3:PL_g2}(b), (c), (d), and (e), corresponding to emitters with ZPL wavelengths 637, 633, 608, and 581 nm, respectively. The extracted $g^{(2)}(0)$ values are below 0.5 for most of the emitters, strongly confirms their single-photon emission characteristics and quantum nature. In the autocorrelation plots, the experimental data are represented by scatter points connected by thin lines, while the corresponding fits are shown as thick solid lines of matching colors. The experimental data is fitted using a three level model given by~\cite{Khaleel2022}. 

\begin{equation}
    \gtwo(\tau) = 1 -(a_1+a_2)\exp(-\frac{|\tau|}{\tau_1})+a_2\exp(-\frac{|\tau|}{\tau_2})
    \label{eq:g2_3lavel}
\end{equation}

where,
\begin{equation}
a_1+a_2=\rho^2(1+A),
\qquad
a_2=\rho^2A,
\end{equation}

Here, $\rho = \rm Signal/(Signal+Background)$, is the signal fraction and $A$ is the bunching amplitudes. The parameter $\tau_1$ denotes the characteristic antibunching time and is typically associated with the radiative lifetime of the emitter, whereas $\tau_2$ corresponds to the characteristic bunching time arising from population dynamics involving a long-lived metastable (shelving) state. The constant term $1$ represents the asymptotic value of the autocorrelation function at long delay times ($|\tau| \rightarrow \infty$), where successive photon emission events become statistically independent. Repeated measurements performed on representative emitters with ZPLs 637, 633, 608, and 581 nm consistently yielded measured $g^{(2)}(0) \leq 0.5$, thereby confirming their single-photon emission characteristics.

A characteristic time constant $\tau_1 \sim 1.11$-$1.8$ ns, is observed for emitters shown in Figure~\ref{fig3:PL_g2}(d,e), corresponding to ZPLs at 608 nm and 581 nm. The fast component is attributed to the excited-state lifetime responsible for antibunching at zero delay. While a slower component, with $\tau_2 \sim 130$–$210$ ns, is associated with population trapping in a metastable state, which temporarily interrupts the emission cycle and gives rise to photon bunching at intermediate delay times. The presence of a clearly separated slow timescale ($\tau_2 \gg \tau_1$) provides strong evidence for a three-level emitter dynamics. In contrast, a simple two-level model would not capture the observed bunching behavior. 

In addition, emitters in Figure~\ref{fig3:PL_g2}(b,c) exhibits a fast antibunching time constant of $\tau_1 \approx 0.42-0.79$ ns. The extracted slow time constant is $\tau_2 $ in ms range; however, this value lies close to the upper limit of the experimentally accessible delay range and should therefore be interpreted with caution. While the fit formally suggests the presence of a slow bunching contribution, the large uncertainty in $\tau_2$ indicates that the metastable-state dynamics are not fully resolved within the present measurement window. Nevertheless, these emitters still exhibit clear antibunching behavior with $g^{(2)}(0) < 0.5$, confirming single-photon emission.

The statistical analysis together with the second-order autocorrelation measurements identified the emitter with a ZPL at 581 nm as the most abundant and highest-purity single-photon emitter among the investigated emitters. Consequently, all stability and power-dependent measurements were performed on this representative emitter. Figure~\ref{fig4:stability}(a) presents the temporal stability measurement acquired over 300 s under continuous-wave excitation. A total of 600 consecutive spectra were recorded with an integration time of 0.5 s per spectrum. The persistent emission line at 581 nm throughout the entire measurement duration demonstrates the excellent photostability and blinking-free nature of the emitter.

To further investigate the emitter performance, power-dependent second-order autocorrelation measurements ($\gtwo(\tau)$), were carried out on the same emitter, as shown in Figure~\ref{fig4:stability}(b). The excitation powers used for each measurement and the corresponding extracted ($\gtwo(\tau=0)$) values are indicated in the respective panels. As the excitation power is increased from 0.5 to 4 mW, the extracted ($\gtwo(\tau=0)$) value gradually increases from (0.15 $\pm$ 0.09) to (0.55 $\pm$ 0.06), as summarized in Figure~\ref{fig4:stability}(c). This increase is likely associated with an enhanced contribution from background fluorescence and/or power-induced dephasing processes at higher excitation powers. Simultaneously, the characteristic antibunching lifetime ($\tau_1$) decreases with increasing excitation power, indicating a faster excitation-emission cycle under stronger optical pumping. The power-dependent photoluminescence intensity of the emitter is shown in Figure~\ref{fig4:stability}(d). Fitting the experimental data (gray circles) with a saturation model (red curve) yields a saturation count rate of 12.23 MHz and a saturation power of 17.49 mW. These results demonstrate that the emitters fabricated using the proposed three-step deterministic fabrication process exhibit excellent photostability, high brightness, and robust single-photon emission characteristics at room temperature.

\section{Conclusion}

These results demonstrate a lithography-free approach for the deterministic generation of single-photon emitters, simplifying fabrication while maintaining a high site-correlated activation yield. In contrast to previous carbon-assisted approaches, which primarily reported emitters centered around 560 nm, our method produces a distinct emitter family dominated by a ZPL at 581 nm with 28\% of all observed emitters.

This study demonstrates the synergistic role of Ga-ion milling, selective carbon seeding, and thermal annealing in the efficient generation of single-photon emitter (SPE) arrays in hBN. Using the proposed three-step fabrication process, approximately 89 of the 100 fabricated sites were realized within or in the immediate vicinity of milled sites in a $(10 \times 10)$ array, corresponding to an emitter yield of $\sim 89\%$. Notably, sharp and photostable SPEs with a zero-phonon-line (ZPL) wavelength centered around 581 nm were found to dominate the emitter population, accounting for approximately $\sim 28\%$ of all observed emitters. In addition, several other prominent emitter groups with ZPL wavelengths near 608, 633, 637, and 660 nm were identified. Second-order correlation measurements unequivocally confirm the quantum nature of these emission centers.

In contrast, a planar hBN flake subjected to identical processing conditions exhibited a significantly lower emitter density than the Ga+C+Annealed sample, highlighting the critical role of Ga-ion-induced defect formation, localized carbon incorporation, and post-fabrication annealing in promoting SPE generation. These findings demonstrate a promising route toward the deterministic and scalable realization of SPE arrays in hBN, with partial control over the emission wavelength. Further optimization of the fabrication parameters is expected to enhance the spectral selectivity and increase the proportion of emitters exhibiting a desired emission wavelength. We anticipate that this approach will provide a robust platform for future investigations of engineered quantum emitters and their integration into quantum photonic technologies.

The deterministic positioning and high activation yield demonstrated here provide a promising platform for integration with nanophotonic cavities and resonant photonic architectures. Such hybrid systems may enable enhanced light--matter interactions, Purcell-enhanced single-photon emission\cite{kumar2022high,singh2023low}, and the exploration of cavity quantum electrodynamics phenomena\cite{kumar2023photonic,sharma2024exceptional}, including the strong-coupling regime\cite{kumar2023universal}, in hBN-based quantum emitters.
\section{Methodology}
\textbf{\emph{Sample preparation flow:---}}
Mechanically exfoliated hBN thin flakes with thickness ~120 nm are transferred onto the cleaned Si/SiO2 substrates (oxide thickness:300 nm). As depicted in Figure~\ref{fig1:sample_prepration}  (a), Ga focused ion beam [Thermoscientific, Helios 5 UC] is employed for milling the nano-grooves with radius ~100 nm and depth ~100 nm. The milling is performed at a 30 kV extraction potential, 40 pA beam current, and stage tilt of 52 degrees. Sequentially selective Carbon deposition is used for filling the nano-grooves with high precision at a stage tilt of 0 degrees using C-beam facility equipped with FIB system. The fabricated samples are subjected to thermal annealing at 850 $ ^ \circ$ C for 1 h in the argon flow at 25 sccm to activate SPEs.

\textbf{\emph{Surface tomography:---}}
Atomic force microscopy [Oxford Instruments Asylum Research Inc, MFP-3D BIO] in tapping mode is employed to obtain the surface profile of the hBN flake after the fabrication of the SPEs. Oxford Asylum Research software is used for data processing. Nano-Grooves with depth of $\sim 39 \nm$ are observed on hBN flakes [see Figure ~\ref{fig1:sample_prepration} (c) and (d)]. The nominal milling depth of 100 \nm and carbon filling of 50 \nm, should result in the milled spots with depth of 50 \nm, however we observed it to be around $\sim 39 \nm$.  The observed variation in the depth could be attributed to the experimental uncertainties in the milling depth and the deposition thickness. 

\textbf{\emph{SPE characterization:---}}
 WITec [Alpha300 R] confocal Raman/PL spectrometer equipped with Hanbury Brown and Twiss (HBT) measurement setup is used for characterizing the fabricated luminescent centers on the hBN. Initially, confocal PL maps are performed to identify the emission spots using 100x objective with NA ~0.9. Upon observing the emission centers, second order correlation measurements are carried out with the same set up in order to confirm single photon nature of the emitters. Excelitas Single-Photon Counting Modules (SPCMs) and Swabian time tagger are employed for time correlation measurements. Clustering analysis of the emitters was performed using the Python-based scikit-learn (sklearn) library. The clustering was carried out on the emitter spectral parameters, including the zero-phonon-line (ZPL) wavelength and full width at half maximum (FWHM), to identify groups of emitters with similar optical characteristics.

 \section*{acknowledgement}
We acknowledge funding support from the National Quantum Mission, an initiative of the Department of Science and Technology, Government of India. We also acknowledge funding from ANRF via grant number SPR/2023/000175. Mangababu acknowledges the IPDF fellowship from IIT Bombay. Parul acknowledges Prime Minister's Research Fellowship. Rohit acknowledges University Grants Commission for research fellowship. We acknowledges the Centre for Sophisticated Instruments and Facilities (CSIF) for access to the FIB-SEM facility and technical support from Mrs. Princy Denis Varghese.

\section*{Author Contributions}

M.A. conceived and designed the study, managed the project, prepared the samples, performed data acquisition and analysis, wrote the original manuscript draft, and supervised the research. R.K. contributed to sample preparation, data acquisition, and optical measurements. B.K. contributed to data analysis, manuscript preparation, and supervision. H.G. contributed to sample preparation. P.S. performed optical characterization of the samples. I.S. contributed to data acquisition. A.K. supervised the project, provided scientific guidance, and acquired funding. All authors discussed the results and contributed to the final version of the manuscript.

\bibliographystyle{apsrev4-2}
\bibliography{ref}
\end{document}